\documentclass[prl,aps,twocolumn]{revtex4}
\usepackage[]{times,amsmath,epsfig,graphicx}

\begin{document}

\title{Generation of a coherent near-infrared Kerr frequency comb in a monolithic microresonator with normal GVD}

\author{Wei Liang, Anatoliy A. Savchenkov, Vladimir S. Ilchenko, Danny Eliyahu, David Seidel, Andrey B. Matsko, and Lute Maleki}

\affiliation{ OEwaves Inc., 465 North Halstead Street, Suite 140, Pasadena, CA 91107}

\begin{abstract}
We demonstrate experimentally, and explain theoretically, generation of a wide, fundamentally phase locked Kerr frequency comb in a nonlinear resonator with a normal group velocity dispersion. A magnesium fluoride whispering gallery resonator characterized with 10~GHz free spectral range and pumped either at 780~nm or 795~nm is used in the experiment. The envelope of the observed frequency comb differs significantly from the Kerr frequency comb spectra reported previously. We show via numerical simulation that, while the frequency comb does not correspond to generation of short optical pulses, the relative phases of the generated harmonics are fixed.
\end{abstract}

\maketitle

A nonlinear monolithic optical resonator pumped with continuous wave (cw) light can produce a broad optical frequency (Kerr) comb \cite{delhaye07n,delhaye08prl,savchenkov08prl,grudinin09ol,razzari10np,ferdous11np,foster11oe,papp11pra,delhaye11prl,liang11ol,savchenkov11np,okawachi11ol,grudinin12oe,johnson12ol,ferdous12oe,li12prl,savchenkov12pra,savchenkov12oe,wang12oe,papp13prx,saha13oe,herr13arch,papp13oe,delhaye13arch}  and a train of ultrashort optical pulses \cite{saha13oe,herr13arch}, generated due to resonant modulation instability effect \cite{nakazawa89jqe,haelterman92ol,coen97prl,serkland99ol,coen01ol,matsko05pra}. The process is phase matched in a broad range of parameters if the group velocity dispersion (GVD) of the resonator modes is anomalous \cite{agha07pra,agha09oe}. However, phase matching is compromised in the case of purely normal GVD. While modulation instability \cite{matsko05pra,haelterman92ol,coen97prl,chembo10pra,matsko11nlo,godey13arch,hansson13pra} as well as mode locking \cite{matsko12ol,coillet13pj} is still possible under this condition, generation of a broad frequency comb has not been previously demonstrated under net normal GVD. Short optical pulses can be created in a Kerr frequency comb system if the optical loss of a nonlinear ring microresonator has specific frequency dependence \cite{wong14spie}. The loss dependence modifies the GVD of the resonator in a way similar to conventional mode locked lasers that can operate at any GVD. This method, though, is not easily utilizable for a large variety of the broadband monolithic microresonators.

In this Letter we report on observation of a stable normal GVD Kerr frequency comb. The shape of the frequency envelope of the comb differs significantly from the earlier predictions and observations \cite{godey13arch,matsko12ol,coillet13pj}. By demodulating the comb on a fast photodiode and observing the phase noise of the generated radio frequency (RF) signal we prove that the frequency harmonics are phase locked \cite{savchenkov08prl}. To ensure generality of the phenomenon we performed the experiment at two different wavelengths, 780~nm and 795~nm, using several different resonators and confirmed generation of the mode locked combs with similar properties.

Using numerical simulations, we reproduce the frequency comb envelope and show that the  comb does not correspond to generation of a short optical pulse in the resonator; rather, the pulses are "dark," i.e. they have lower power as compared to the DC background in the resonator. The frequency comb produces bright pulses at the resonator output, due to the interference with the pump light. It is worth noting that generation of bright pulses inside the resonator usually leads to generation of "dark" pulses at the resonator output \cite{matsko13oe}. The pump light has to be filtered out to enable observation of the bright pulses \cite{herr13arch}.
\begin{figure}[htb]
\centerline{\includegraphics[width=8.cm]{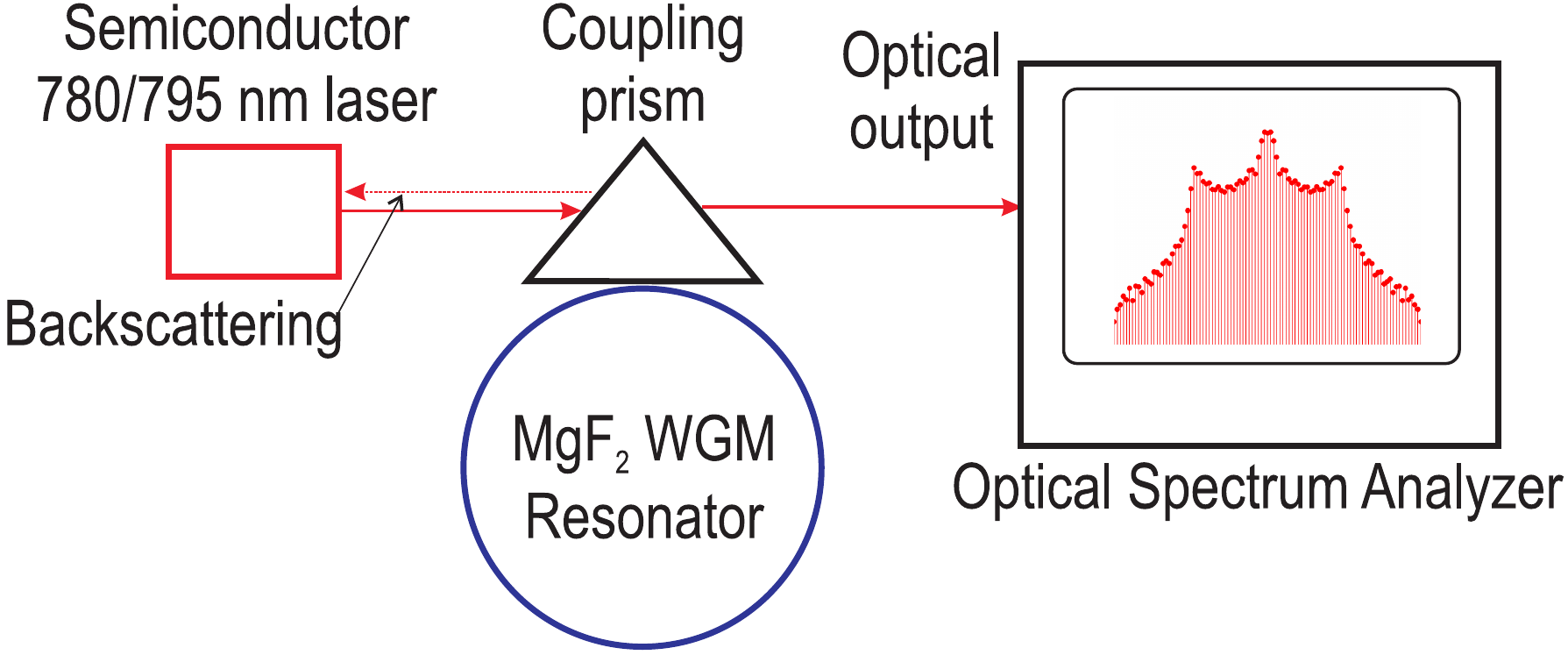}}
\caption{{\small Schematic of the experiment. Either a 780~nm or 795~nm distributed feedback semiconductor laser is self-injection locked to a magnesium fluoride whispering gallery mode resonator. An optical frequency comb is generated in the resonator and analyzed using an optical spectrum analyzer. The resonator is characterized with essentially normal GVD and the shape of comb envelope has three distinct maxima. }\label{figure1}}
\end{figure}

Numerical simulations are particularly useful for validation of the experimental results. It has been argued previously that there are multiple reasons for comb generation in a multimode nonlinear resonator characterized with normal GVD. For example, avoided crossing between resonator modes allow achieving anomalous GVD locally, which results in comb generation \cite{savchenkov12oe}. A resonator can have both normal and anomalous GVD mode families, so the frequency comb is preferably generated involving the anomalous GVD modes \cite{savchenkov11np}. Our simulation reported in this Letter reproduces the observed comb envelope for realistic experimental parameters. It confirms that the comb can be generated in the continuously pumped normal GVD nonlinear resonator without any additional technique to achieve phase matching. There is no need for engineering morphology of the resonator \cite{savchenkov11np} to introduce a mode family with a specific geometrical dispersion to ensure local phase matching of the nonlinear process. While this kind of shape adjusting is possible, it is rather complex.

These results are important since they pave the way toward generation of a broad Kerr frequency comb at any desirable wavelength, including visible and ultraviolet, in a nonlinear monolithic resonator of any morphology. The normal GVD combs are suitable for interrogation of various atomic transitions and for creation of compact optical clocks \cite{savchenkov13ol}.

In the experiment we use a magnesium fluoride (MgF$_2$) whispering gallery mode (WGM) resonators with approximately 7~mm diameter, which corresponds to $9.9$~GHz free spectral range (FSR) at 780~nm as well as 795~nm cw pumping wavelength. The resonators are fabricated from a commercially available z-cut MgF$_2$ optical window by mechanical polishing. The shape of the resonators in the vicinity of mode localization is that of an oblate spheroid with removed polar caps for easier integration. The resonator modes have loaded quality factor exceeding $Q=2.5\times 10^9$, which corresponds to half width at half maximum (HWHM)  $\gamma=77$~kHz. The bandwidth is worse as compared to HWHM of $15-20$~kHz observed at 1550~nm with the resonators, caused by the increase of Rayleigh scattering at shorter wavelength.

The light emitted from a distributed feedback (DFB) semiconductor laser is coupled to the resonator with a coupling prism. The backscattered light from the resonator is used to lock the laser to the WGM of interest \cite{liang10ol}.  The light exiting the prism in the forward direction is collimated into a single mode fiber and analyzed with an optical spectrum analyzer. The optical power emitted by the laser is approximately 10~mW, and about 60\% of the power is lost due to imperfect mode matching. The threshold of the frequency comb generation is approximately 0.3~mW.

\begin{figure}[htb]
\centerline{\includegraphics[width=6.5cm]{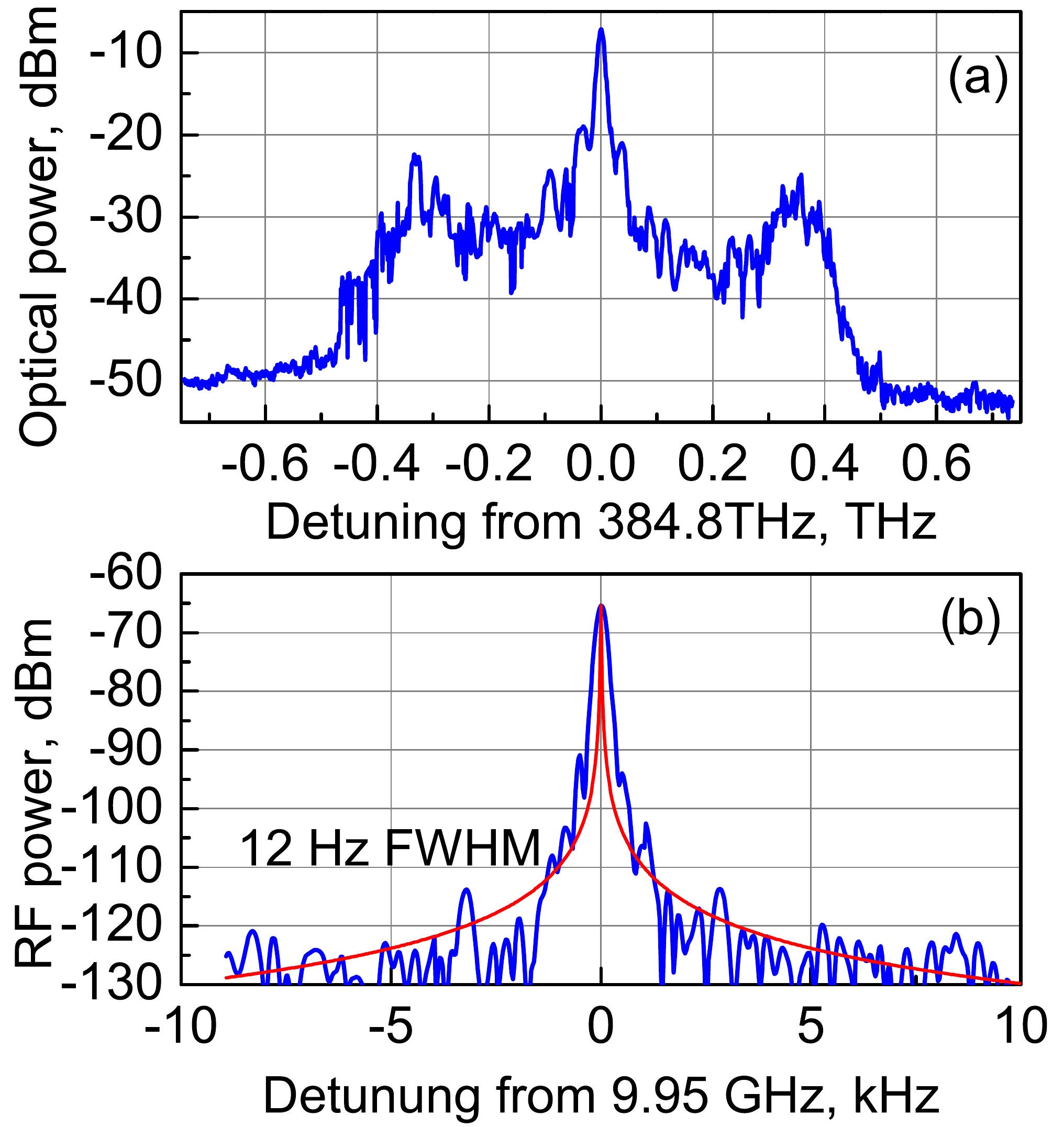}}
\caption{{\small A fundamental optical frequency comb generated with the normal GVD MgF$_2$ WGM resonator, (a), and RF frequency signal generated at the photodiode by the comb (20~kHz resolution bandwidth (RBW)), (b). The comb lines separated by 10~GHz are not resolved by the optical spectrum analyzer. The red line is the Lorentzian fit of the skirts of the RF signal. Such a fit is a good measure of the instantaneous linewidth of the RF radiation.  }\label{figure2}}
\end{figure}

The spectrum of the optical frequency comb observed at 780~nm is shown in Fig.~(\ref{figure2}a). Our optical spectrum analyzer did not have enough resolution to separate the comb lines, so we were able to see the envelope only. To verify the comb repetition rate we sent the optical signal to a fast photodiode, and then to an RF spectrum analyzer that revealed a spectrally pure beat note at 9.9~GHz Fig.~(\ref{figure2}b). The measurement confirmed that the comb repetition rate is the fundamental one, i.e. it coincides with the FSR of the resonator. The comb has approximately 80 harmonics, according to the envelope width measurement.

The spectrum of the RF signal is helpful for estimating the contrast of the comb frequency harmonics. The RF spectrum analyzer measures the power spectrum of the photocurrent, proportional to $i_{ph}^2$. We estimate the RF power generated at the comb repetition frequency to be proportional to $P_{pump} P_{sideband}$, where $P_{sideband}/ P_{pump} \approx 10^{-2}$. Let us assume that the optical comb is overlapped with white noise having power density $S_{noise}$. Mixing the noise with the pump light results in an RF signal with power $P_{pump} S_{noise} RBW$.  The measured RF spectrum has at least 50~dB contrast with no spurious RF signal visible. This means that $S_{noise}<10^{-5}P_{sideband}/RBW \simeq -110$~dBm$/$Hz and the optical contrast is at least 80~dB, when a proper measurement with 1~Hz RBW is performed.

\begin{figure}[htb]
\centerline{\includegraphics[width=6.5cm]{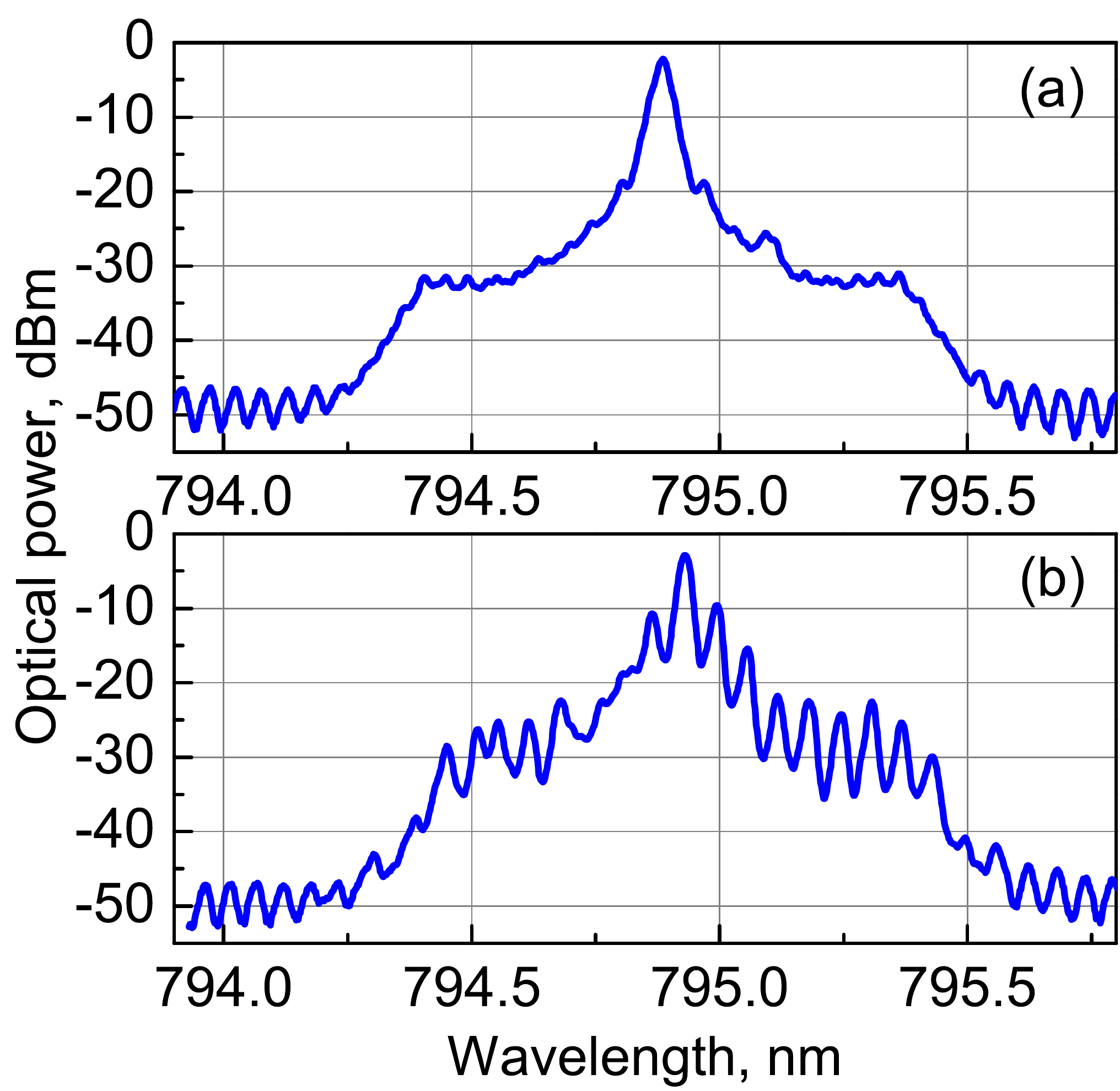}}
\caption{{\small Optical spectra of Kerr frequency combs observed at 795~nm. (a) A nearly ideal comb envelope with slightly pronounced dominance of forth-order harmonics (seen at 40~GHz from the pump frequency). (b) The comb with dominating third-order harmonics. Both comb envelopes show some irregularities resulting from the mode interaction in the multi-mode resonator. }\label{figure2p5}}
\end{figure}

We observed similar comb envelope at 795~nm Fig.~(\ref{figure2p5}). The threshold cw power was approximately the same as one in the experiment with  780~nm light. The comb generated $9.84$~GHz RF signal with a high speed photodiode. The signal had reasonably good spectral purity characterized with single sideband phase noise of $-106$~dBc/Hz at 10~kHz frequency offset, which proves phase locking of the comb harmonics.

The observed normal GVD comb had several stable branches, similarly to the case of anomalous-GVD frequency combs, as illustrated by Fig.~(\ref{figure2p5}). While the branches were characterized with similar spectral purity, they had different power distribution among the harmonics within the comb envelope. Generation of chaotic signals was also observed. In addition to the multitude of the stable solutions, we observed the influence of resonator mode interaction on the comb generation \cite{savchenkov12oe}. To reduce the influence of the mode interaction regime we tuned the resonator spectrum by changing its temperature.

The shape of the envelope of the frequency comb looks rather different from the conventional triangular comb envelope shapes predicted and observed in resonators with normal GVD \cite{godey13arch,matsko12ol,coillet13pj}. The envelope has symmetric peak-like structures followed by steep skirts. The "peaks" of the observed comb envelope are located at 35$^{th}$ mode from the carrier. We examined the observation using numerical simulation and found out that the given shape of the frequency comb may indeed be generated in the resonator.

We need to know the value of GVD to perform the simulation. Using Sellmeier equation for MgF$_2$ and an asymptotic expression describing the spectrum of a dielectric spherical resonator we find refractive index of the material, $n_0=1.38$, and normal GVD, $\beta_2 \simeq 21.8$~ps$^2/$km at the 780~nm pump wavelength. The GVD drops by 3\% if 795~nm pump is selected instead. It is worth noting that MgF$_2$ is characterized  with anomalous GVD at longer wavelengths and zero GVD point is located approximately at 1,377~nm for the resonator of given dimensions. The GVD value corresponds to 2.96~kHz frequency difference between adjacent FSRs ($2\nu_{0}-\nu_+-\nu_- \simeq 2.96$~kHz, where $\nu_0$ is the linear frequency of the pumping light, $\nu_\pm$ are the frequencies of the optical sidebands, FSR is given by $\nu_+-\nu_0\simeq \nu_0-\nu_-\simeq 9.96$~GHz). The dimensionless GVD parameter, defined as $D=(2\nu_{0}-\nu_+-\nu_-)/\gamma$ \cite{matsko05pra}, is equal to $0.039$ for this resonator. MgF$_2$ is characterized with cubic nonlinearity $n_2=0.9\times 10^{-16}$~cm$^2/$W \cite{milam77apl}. The mode volume is ${\cal V} \approx 10^{-6}$~cm$^3$.

We numerically solve a set of ordinary differential equations describing the behavior of 101 optical modes, expecting all the modes to be identical and completely overlapping in space.
\begin{equation} \label{set}
\dot{\hat a}_j=-(2\pi \gamma+i\omega_j)\hat a_j+ \frac{i}{\hbar} [\hat V,\hat a_j]+F_0 e^{-i\omega t} \delta_{j0,j},
\end{equation}
where ${\hat a}_j$ is an annihilation operator for $j^{th}$ mode, $\delta_{j0,j}$ is the Kronecker's delta, $\hat V= -(\hbar g/2) (\hat e^\dag)^2 \hat e^2$, $\hat e= \sum \hat a_j$. We introduce a coupling constant $g=\hbar \omega_0^2 c n_2/({\cal V}n_0^2)=6\times10^{-4}$~s$^{-1}$ as well as a dimensionless pumping constant $f=(F_0/2\pi \gamma)(g/2\pi \gamma)^{1/2}$, where $F_0=(4 \pi \gamma P/(\hbar \omega_0))^{1/2}$ stands for the amplitude of the continuous wave external pump, $\omega_0=2 \pi \nu_0$, and $P$ is the pump power. The pumping constant is $f=7$ for $P=2.2$~mW. It is also useful to introduce $P_0=45~\mu$W describing the lowest threshold of hyper-parametric oscillation for optimal anomalous GVD value. The threshold is higher for the case of anomalous GVD.

The external continuous wave pump is applied to the central mode of the mode group ($j=j_0$). Only the second order frequency dispersion, as described by parameter $D$, is taken into account. The only free parameters are the pump amplitude, $f$, and frequency detuning, $\Delta=(\nu-\nu_0)/\gamma,$ where $\nu$ is the frequency of the optical pump.
\begin{figure}[htb]
\centerline{\includegraphics[width=7.5cm]{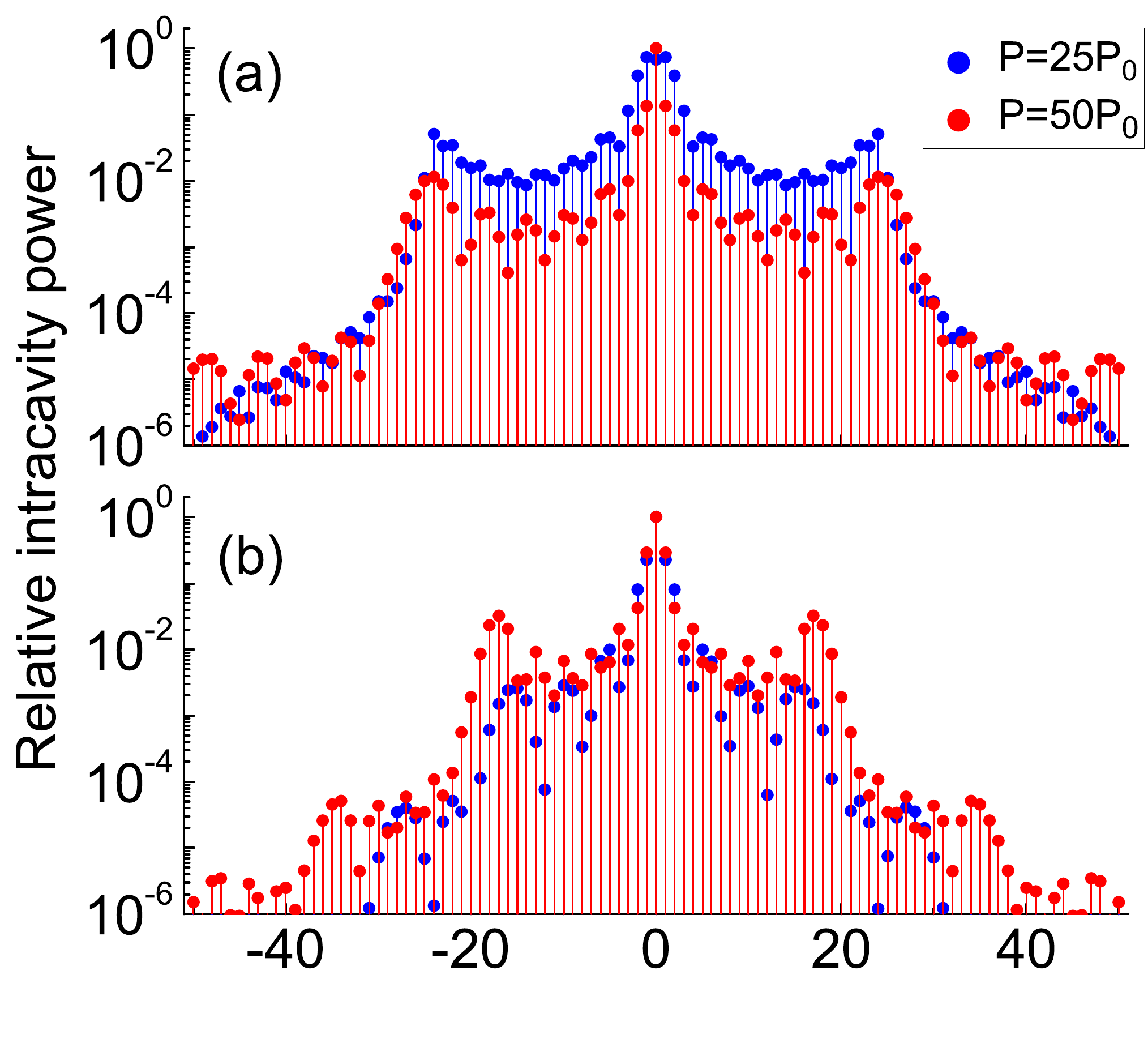}}
\caption{{\small Numerically simulated spectra of optical frequency combs for two different pump power levels and GVD parameters. (a) $D=0.039$, (b) $D=0.078$. }\label{figure3}}
\end{figure}

\begin{figure}[htb]
\centerline{\includegraphics[width=7.5cm]{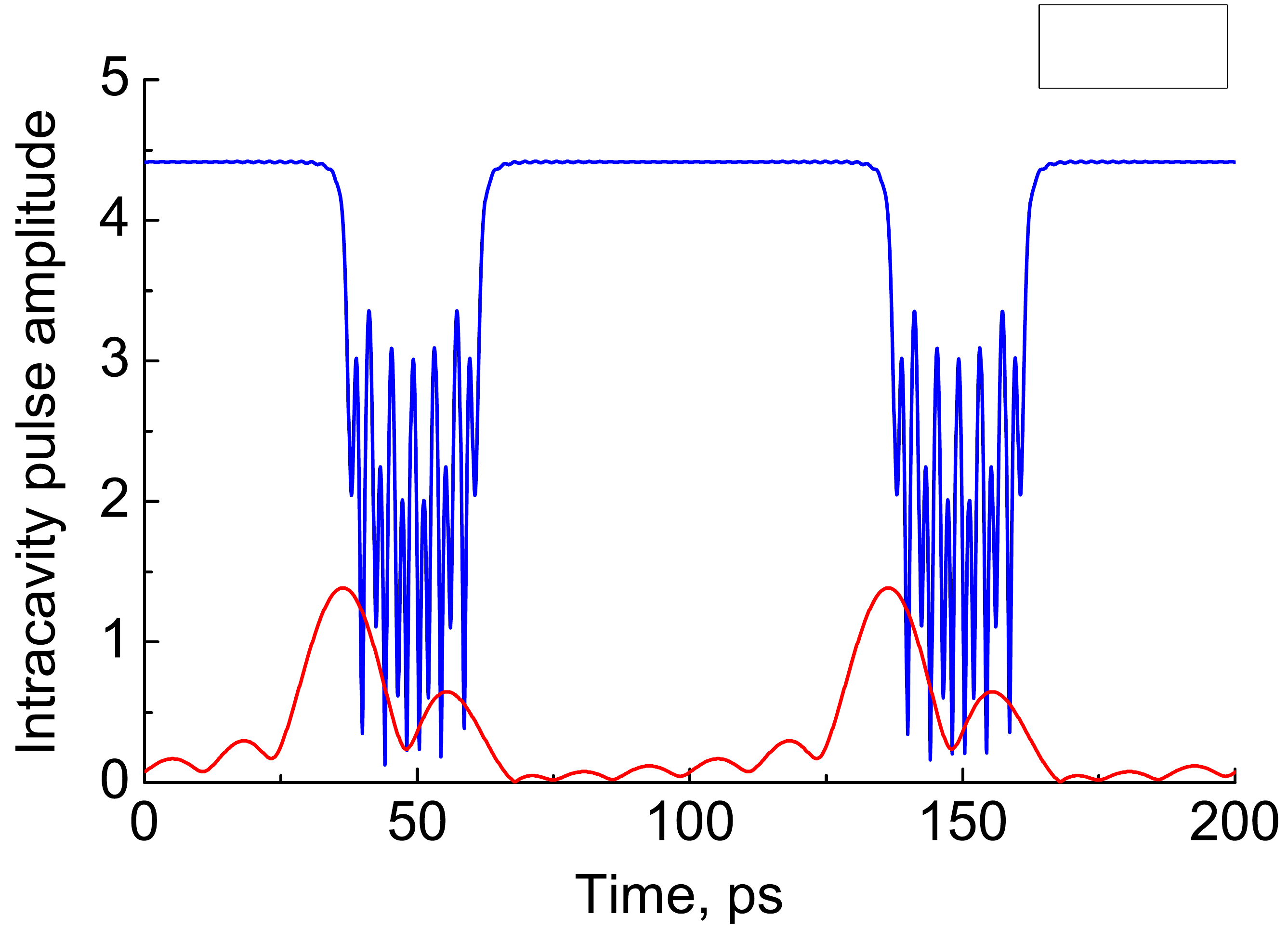}}
\caption{{\small Numerically simulated envelope of intracavity optical pulses in terms of normalized amplitude $|A|(g/\gamma)^{1/2}$ ($j=1\dots 101$, blue line) and the pulse formed by only a limited number of modes with no pump frequency included ($j=40\dots 47$, red line). }\label{figure4}}
\end{figure}

Results of the simulation are presented in Fig.~(\ref{figure3}) and Fig.~(\ref{figure4}). The simulation show that it is possible to generate a stable frequency comb in a normal GVD resonator. The simulated envelope resembles the frequency comb envelope observed experimentally (compare Fig.~(\ref{figure2}), Fig.~(\ref{figure2}a), and Fig.~(\ref{figure3})). The position of the "peaks" at the envelope depends on the normalized GVD value (parameter $D$) and does not depend much on the pump power. The mode where the "peak" occurs has the approximate number $|j-j_0|\approx D^{-1}$.

To generate a broader frequency comb one needs to reduce $D$. This parameter depends on the GVD of the resonator as well as loading of the resonator modes. It means, in particular, that in the experiment the comb spectrum should drastically change depending on the degree of resonator loading as $D$ decreases with the loading increase. This is what was observed in our measurements.

To verify phase locking of the comb harmonics we found the envelope of the pulse generated in the resonator (Fig.~\ref{figure4}). The frequency comb corresponds to a manifold of short "dark" pulses travelling inside the resonator. The pulse envelope does not change in time which proves mode locking (and phase locking) of the comb harmonics. Bright high contrast pulses can be created if only several frequency harmonics are selected ($j=40-47$) and the carrier harmonic is excluded. The selection procedure can be realized by retrieving the frequency comb out of the resonator using the second evanescent field coupler and by filtering the output with a bandpass filter. This procedure, though, results in the expected increase in the pulse duration.

In conclusion, we have demonstrated experimentally the feasibility of generation of a phase locked optical frequency comb in a nonlinear resonator characterized with normal GVD. The optical envelope of the comb was recorded and phase locking was confirmed via measurement of the RF signal generated by the comb with a fast photodiode. The possibility of comb generation was confirmed by numerical simulation.

The authors acknowledge support from Defense Sciences Office of Defense Advanced Research Projects Agency under contract No. W911QX-12-C-0067 as well as support from Air Force Office of Scientific Research under contract No. FA9550-12-C-0068.


\begin{thebibliography}{99}

\bibitem{delhaye07n} P. Del-Haye, A. Schliesser, O. Arcizet, T. Wilken, R. Holzwarth, and T. J. Kippenberg, {\lq\lq} Optical frequency comb generation from a monolithic microresonator," Nature {\bf 450}, 1214--1217 (2007).

\bibitem{delhaye08prl} P. Del-Haye, O. Arcizet, A. Schliesser, R. Holzwarth, and T. J. Kippenberg, {\lq\lq}Full stabilization of a microresonator-based optical frequency comb," Phys. Rev. Lett. {\bf 101}, 053903 (2008).

\bibitem{savchenkov08prl} A. A. Savchenkov, A. B. Matsko, V. S. Ilchenko, I. Solomatine, D. Seidel, and L. Maleki, {\lq\lq}Tunable optical frequency comb with a crystalline whispering gallery mode resonator," Phys. Rev. Lett. {\bf 101}, 093902 (2008).

\bibitem{grudinin09ol} I. S. Grudinin, N. Yu, and L. Maleki, {\lq\lq}Generation of optical frequency combs with a CaF$_2$ resonator," Opt. Lett. {\bf 34}, 878--880 (2009).

\bibitem{razzari10np} L. Razzari, D. Duchesne, M. Ferrera, R. Morandotti, S. Chu, B. E. Little, and D. J. Moss, {\lq\lq}CMOS-compatible integrated optical hyper-parametric oscillator," Nature Photonics {\bf 4}, 41--45 (2010).

\bibitem{ferdous11np} F. Ferdous, H. X. Miao, D. E. Leaird, K. Srinivasan, J. Wang, L. Chen, L. T. Varghese, and A. M. Weiner, {\lq\lq}Spectral line-by-line pulse shaping of on-chip microresonator frequency combs," Nature Photonics {\bf 5}, 770--776 (2011).

\bibitem{foster11oe} M. A. Foster, J. S. Levy, O. Kuzucu, K. Saha, M. Lipson, and A. L. Gaeta, {\lq\lq}Silicon-based monolithic optical frequency comb source," Optics Express {\bf 19}, 14233--14239 (2011).

\bibitem{papp11pra} S. B. Papp and S. A. Diddams, {\lq\lq}Spectral and temporal characterization of a fused quartz microresonator optical frequency comb," Phys. Rev. A {\bf 84}, 053833 (2011).

\bibitem{delhaye11prl} P. Del-Haye, T. Herr, E. Gavartin, M. L. Gorodetsky, R. Holzwarth, and T. J. Kippenberg, {\lq\lq}Octave spanning tunable frequency comb from a microresonator," Phys. Rev. Lett. {\bf 107} 063901 (2011).

\bibitem{liang11ol} W. Liang, A. A. Savchenkov, A. B. Matsko, V. S. Ilchenko, D. Seidel, and L. Maleki, {\lq\lq}Generation of near-infrared frequency combs from a MgF$_2$ whispering gallery mode resonator," Opt. Lett. {\bf 36}, 2290--2292 (2011).

\bibitem{savchenkov11np} A. A. Savchenkov, A. B. Matsko, W. Liang, V. S. Ilchenko, D. Seidel, and L. Maleki, {\lq\lq}Kerr combs with selectable central frequency," Nature Photon. {\bf 5}, 293--296 (2011).

\bibitem{okawachi11ol} Y. Okawachi, K. Saha, J. S. Levy, Y. H. Wen, M. Lipson, and A. L. Gaeta, {\lq\lq}Octave spanning frequency comb generation in a silicon nitride chip," Opt. Lett. {\bf 36}, 3398--3400 (2011).

\bibitem{grudinin12oe} I. S. Grudinin, L. Baumgartel, and N. Yu, {\lq\lq}Frequency comb from a microresonator with engineered spectrum," Opt. Express {\bf 20}, 6604--6609 (2012).

\bibitem{johnson12ol} A. R. Johnson, Y. Okawachi, J. S. Levy, J. Cardenas, K. Saha, M. Lipson, and A. L. Gaeta, {\lq\lq}Chip-based frequency combs with sub-100 GHz repetition rates," Opt. Lett. {\bf 37}, 875--877 (2012).

\bibitem{ferdous12oe} F. Ferdous, H. X. Miao, P. H. Wang, D. E. Leaird, K. Srinivasan, L. Chen, V. Aksyuk, and A. M. Weiner, {\lq\lq}Probing coherence in microcavity frequency combs via optical pulse shaping," Opt. Express {\bf 20}, 21033- -21043 (2012).

\bibitem{li12prl} J. Li, H. Lee, T. Chen, and K. J. Vahala, {\lq\lq}Low-pump-power, low-phase-noise, and microwave to millimeter-wave repetition rate operation in microcombs," Phys. Rev. Lett. {\bf 109}, 233901 (2012).

\bibitem{savchenkov12pra} A. A. Savchenkov, A. B. Matsko, W. Liang, V. S. Ilchenko, D. Seidel, and L. Maleki, {\lq\lq}Transient regime of Kerr-frequency-comb formation," Phys. Rev. A  {\bf 86}, 013838 (2012).

\bibitem{savchenkov12oe} A. A. Savchenkov, A. B. Matsko, W. Liang, V. S. Ilchenko, D. Seidel, and L. Maleki, {\lq\lq}Kerr frequency comb generation in overmoded resonators," Opt. Express {\bf 20}, 27290--27298 (2012).

\bibitem{wang12oe} P. H. Wang, F. Ferdous, H. X. Miao, J. Wang, D. E. Leaird, K. Srinivasan, L. Chen, V. Aksyuk, and A. M. Weiner, {\lq\lq}Observation of correlation between route to formation, coherence, noise, and communication performance of Kerr combs," Opt. Express {\bf 20}, 29284--29295 (2012).

\bibitem{papp13prx} S. B. Papp, P. Del�Haye, and S. A. Diddams, {\lq\lq}Mechanical control of a microrod-resonator optical frequency comb," Phys. Rev. X {\bf 3}, 031003 (2013).

\bibitem{saha13oe} K. Saha, Y. Okawachi, B. Shim, J. S. Levy, R. Salem, A. R. Johnson, M. A. Foster, M. R. E. Lamont, M. Lipson, and A. L. Gaeta, {\lq\lq}Modelocking and femtosecond pulse generation in chip-based frequency combs," Opt. Express {\bf 21}, 1335--1343 (2013).

\bibitem{herr13arch} T. Herr, V. Brasch, J. D. Jost, C. Y. Wang, N. M. Kondratiev, M. L. Gorodetsky, and T. J. Kippenberg, {\lq\lq}Mode-locking in an optical microresonator via soliton formation," arXiv:1211.0733v2 (2013).

\bibitem{papp13oe} S. B. Papp, P. Del�Haye, and S. A. Diddams, {\lq\lq}Parametric seeding of a microresonator optical frequency comb," Opt. Express {\bf 21}, 17615--17624 (2013).

\bibitem{delhaye13arch} P. Del'Haye, S. B. Papp, and S. A. Diddams, {\lq\lq}Self-injection locking and phase-locked states in microresonator-based optical frequency combs," arXiv:1307.4091 (2013).

\bibitem{nakazawa89jqe} M. Nakazawa, K. Suzuki, and H. A. Haus, {\lq\lq}The modulational instability laser-Part I: Experiment," IEEE J. Quantum Electron. {\bf 25}, 2036--2044 (1989).

\bibitem{haelterman92ol} M. Haelterman, S. Trillo, and S. Wabnitz, {\lq\lq}Additive-modulation-instability ring laser in the normal dispersion regime of a fiber," Opt. Lett. {\bf 17}, 745--747 (1992).

\bibitem{coen97prl} S. Coen and M. Haelterman, {\lq\lq}Modulational instability induced by cavity boundary conditions in a normally dispersive optical fiber," Phys. Rev. Lett. {\bf 79}, 4139--4142 (1997).

\bibitem{serkland99ol} D. K. Serkland and P. Kumar, {\lq\lq}Tunable fiber-optic parametric oscillator," Opt. Lett. {\bf 24}, 92--94 (1999).

\bibitem{coen01ol} S. Coen and M. Haelterman, {\lq\lq}Continuous-wave ultrahigh-repetition-rate pulse-train generation through modulational instability in a passive fiber cavity," Opt. Lett. {\bf 26}, 39--41 (2001).

\bibitem{matsko05pra} A. B. Matsko, A. A. Savchenkov, D. Strekalov, V. S. Ilchenko, and L. Maleki, {\lq\lq}Optical hyperparametric oscillations in a whispering-gallery-mode resonator: Threshold and phase diffusion," Phys. Rev. A {\bf 71}, 033804 (2005).

\bibitem{agha07pra} I. H. Agha, Y. Okawachi, M. A. Foster, J. E. Sharping, and A. L. Gaeta, {\lq\lq}Four-wave-mixing parametric oscillations in dispersion-compensated high-Q optical microspheres," Phys. Rev. A {\bf 76}, 043837 (2007).

\bibitem{agha09oe} I. H. Agha, Y. Okawachi, and A. L. Gaeta, {\lq\lq}Theoretical and experimental investigation of broadband cascaded four-wave mixing in high-Q microspheres," Opt. Express {\bf 17}, 16209-16215 (2009).

\bibitem{hansson13pra} T. Hansson, D. Modotto, and S. Wabnitz, {\lq\lq}Dynamics of the modulational instability in microresonator frequency combs," Phys. Rev. A {\bf 88}, 023819 (2013).

\bibitem{godey13arch} C. Godey,  I. Balakireva, A.Coillet, and Y. K. Chembo, {\lq\lq}Stability analysis of the Lugiato-Lefever model for Kerr optical frequency combs. Part I: case of normal dispersion,"  arXiv:1308.2539 (2013).

\bibitem{chembo10pra} Y. K. Chembo and N. Yu, {\lq\lq}Modal expansion approach to optical-frequency-comb generation with monolithic whispering-gallery-mode resonators," Phys. Rev. A {\bf 82}, 033801 (2010).

\bibitem{matsko11nlo} A. Matsko, A. Savchenkov, W. Liang, V. Ilchenko, D. Seidel, and L. Maleki, {\lq\lq}Group Velocity Dispersion and Stability of Resonant Hyper-Parametric Oscillations," in Nonlinear Optics: Materials, Fundamentals and Applications, OSA Technical Digest (CD) (Optical Society of America, 2011), paper NWD2.

\bibitem{matsko12ol} A. B. Matsko, A. A. Savchenkov, and L. Maleki, "Normal group-velocity dispersion Kerr frequency comb," Opt. Lett. {\bf 37}, 43-45 (2012)

\bibitem{coillet13pj} A. Coillet, I. Balakireva, R. Henriet, K. Saleh, L. Larger, J. M. Dudley, C. R. Menyuk, and Y. K. Chembo, {\lq\lq}Azimuthal Turing patterns, bright and dark cavity solitons in Kerr combs generated with whispering-gallery-mode resonators," IEEE Photonics J. {\bf 5}, 6100409 (2013).

\bibitem{wong14spie} C. W. Wong, S.-W. Huang, S. Combrie, P. Colman, A. De Rossi, C. A. Husko, L. Maleki, A. B. Matsko, J. F. McMillan, J. Yang, and H. Zhou, {\lq\lq}Chip-scale ultrafast solitons and frequency comb mode-locking,” SPIE Photonics West 2014, Paper 8960-2 (2014).

\bibitem{matsko13oe} A. B. Matsko and L. Maleki, {\lq\lq}On timing jitter of mode locked Kerr frequency combs," Opt. Express {\bf 21}, 28862-28876 (2013).

\bibitem{savchenkov13ol} A. A. Savchenkov, D. Eliyahu, W. Liang, V. S. Ilchenko, J. Byrd, A. B. Matsko, D. Seidel, and L. Maleki, {\lq\lq}Stabilization of a Kerr frequency comb oscillator," Opt. Lett. {\bf 38}, 2636-2639 (2013).

\bibitem{liang10ol} W. Liang, V. S. Ilchenko, A. A. Savchenkov, A. B. Matsko, D. Seidel, and L. Maleki, {\lq\lq}Whispering-gallery-mode-resonator-based ultranarrow linewidth external-cavity semiconductor laser," Opt. Lett. {\bf 35}, 2822-2824 (2010).

\bibitem{milam77apl} D. Milam, M. J. Weber, and A. J. Glass, {\lq\lq} Nonlinear refractive index of fluoride crystals,” Appl. Phys. Lett. {\bf 31}, 822-824 (1977).

\end{thebibliography}
\end{document}